\documentclass[a4paper,12pt]{article}
\usepackage{graphicx}

\begin{document}

\centerline{\bf ORBIT OF BINARY 15 MONOCEROTIS}
\medskip
\centerline{Zorica Cvetkovi\'c, I\v stvan Vince and Slobodan
Ninkovi\'c}
\medskip
\centerline{\it Astronomical Observatory, Volgina 7, 11060
Belgrade 38, Serbia}
\medskip
\centerline{Email: zcvetkovic@aob.bg.ac.yu}

\bigskip
\bigskip

\small{\bf Abstract:} In the paper new orbital elements obtained
from speckle interferometric measurements only for 15 Mon are
reported. For this binary they were determined earlier by
combining spectroscopic and speckle measurements and spectroscopic
and astrometric measurements. A new speckle measurement dating
after the periastron passage of the secondary has a significant
discrepancy from the ephemeridal value. Our revised orbit has a
period significantly longer than the earlier ones. With regard
that in the case of the earlier orbits the period was determined
from spectroscopic measurements, i.e. from radial velocity data,
the existing data of such kind for this binary are analyzed here
and combined with our orbital elements. The first speckle
measurement originates from 1988 and up to now three more speckle
measurements have been made. This pair is bright (apparent
magnitude of primary 4.66, i.e. 5.9 of secondary), of an early
spectral type (primary O7V, secondary O9.5Vn) and is a member of
open cluster NGC 2264. The distance to this cluster has been
determined many times and different values have been reported. By
analyzing the available data we find that the distance to 15 Mon
is most likely about 750 pc, i.e. the total mass most likely about
53.4${\cal M}_{\odot}$. The new orbital elements combined with
this distance yield a total mass expected for the spectral types
of the components.

\medskip{\bf Keywords:} binaries: visual -- binaries: spectroscopic
-- stars: individual (15 Mon, HD 47839) -- fundamental parameters
(apparent and absolute magnitudes, masses, parallaxes, spectral
types).

\section{Introduction}

The bright O star 15 Monocerotis (15 Mon) in the Washington Double
Star Catalog (WDS) (Mason et al. 2006) has a designation
WDS06410+0954, whereas that of the discoverer is CHR 168 Aa. The
identification numbers in other catalog are HIP 31978, HR 2456,
HD 47839 and 1725 in The ninth catalog of spectroscopic binary
orbits (SB9). According to WDS the primary has a visual apparent
magnitude of 4.66 and spectrum O7Ve, whereas for the secondary one
finds in the literature (Gies et al. 1993) 5.90 and O9.5Vn,
respectively. In the Hipparcos Catalog (ESA 1997) a trigonometric
parallax of 3.19$\pm$0.73 mas is given. The first preliminary
orbital elements were determined by combining the speckle and
spectroscopic measurements (Gies et al. 1993). A few years
afterwards new radial velocity measurements and the first
astrometric one formed a base to a minor revision of these
elements (Gies et al. 1997).

On the basis of published mean radial velocities over the period
1902 - 1993 Gies et al. (1993) determined the spectroscopic
orbital elements for 15 Mon and the remaining ones using the
speckle measurements (Table 1. - first orbit). The companion in
this pair was firstly detected from the speckle measurements in
1988 and by 1993 two more speckle measurements were made. During
those five years the position angle changed (increased) by $
24^{\circ}$, whereas the separation decreased from 57 mas to 40
mas. Assuming the values for $P$, $T$, $e$ and $\omega$ obtained
from the spectroscopic measurements the ones for $a$, $i$ and
$\Omega$ were obtained from the speckle measurements. So Gies et
al. (1993) find a period of 25.2 years. The same authors assumed
that this binary is at a distance of 950 pc, a value found by
Perez et al. (1987) for the distance of NGC 2264. In this way it
became possible to determine the masses of the components of this
early spectral type binary: 34${\cal M}_{\odot}$ for the primary
and 19${\cal M}_{\odot}$ for the secondary. In the given paper the
authors say that the spectroscopic elements are defined
essentially on the basis of four measurements with a large
velocity. If these measurements were eliminated, the other
velocity values would show no variations and it would be
impossible to notice the periodicity. Besides, there is a large
gap in the radial velocity data for the period 1923 - 1969 so that
the authors admit that the orbital period can be even twice as
long.

After four years, based on new radial velocity measurements and
the first astrometric one of the massive binary 15 Mon made with
the Hubble Space Telescope Fine Guidance Sensors (HST FGS), Gies
et al. (1997) published new orbital elements combining
spectroscopic and astrometric data (Table 1. - second orbit).
These measurements confirmed that the companion is very near the
periastron. They obtained from the astrometric data for the epoch
1996.0742 a separation of 22.3-25.1 mas, i. e. 115.9-117.9 degrees
for the position angle depending on the assumed magnitude
difference. These values for both separation and position angle
exceed those following from the orbit calculated before. Hence
this measurement could yield no significant improvement in the
orbit so that it was revised largely on the basis of new radial
velocity measurements performed between 1986 and 1996 and earlier
ones after 1969. The new value for the period found then was 23.6
years. The orbital elements (Gies et al. 1997) are in the Sixth
Catalog of Orbits of Visual Binary Stars (Hartkopf \& Mason 2006)
and in The ninth catalog of spectroscopic binary orbits
(Pourbaix et al. 2004). After assuming a distance of 950 pc and
combining it with the new orbital elements the authors mentioned
above found masses of 35${\cal M}_{\odot}$ and 24${\cal
M}_{\odot}$ for the primary and secondary, respectively.

The distance to open cluster NGC 2264, in which 15 Mon is a
member, has been determined many times. In the article by Perez et
al. (1987) one finds about twenty cited references concerning this
distance determination. From these values a mean value equal to
$d$=735$\pm$106 pc was derived. Also using photometric and
spectroscopic measurements Perez et al. found that NGC 2264 is at
$d$=950$\pm$75 pc from the Sun. In the more recent literature we
find that for the distance to NGC 2264 values between 700 pc and
800 pc have been largely proposed (Ramirez et al. 2004, Dahm \&
Simon 2005, Kharchenko et al. 2005, Dahm et al. 2007, etc).
Consequently, the distance of 950 pc seems to exceed the true one.
 On the other hand, the trigonometric parallax given in Hipparcos
appears to be unrealistic because a too small value of only 313 pc
results. Experience has shown that the HIPPARCOS parallaxes of
order of a few mas are very unreliable.

%
%U radu (Mason et al. 1998) nalazimo vrednost od 720 pc, u (Ramirez
%et al. 2004) 776 pc, u (... AJ 129, 2005) 760-800 pc, u
%(Kharchenko et al. 2005) 660 pc, u (... AJ 134, 2007) 800 pc, itd.

\section{Orbital Elements from the Speckle Measurements}

In the Fourth Catalog of Interferometric Measurements of Binary
Stars (Hartkopf et al. 2006) we find another measurement from the
epoch 2001.0197 made by Mason and coworkers which took place after
the periastron (when due to the small separation the components
could not be resolved by using speckle interferometry). From the
preceding measurement from 1993 the change in the position angle
has attained a value of about $200^{\circ}$, whereas the
separation has increased to 55 mas. The residuals from the
ephemeridal values are large, $-108.9$ degrees in the position
angle and 16 mas in the separation (Table 2). It should be borne
in mind that the earlier orbital elements were derived from
measurements covering a short arc of the orbit and from uncertain
radial velocities. The last speckle measurement enlarges
significantly the observed arc of the orbit and makes possible a
more reliable determination of the orbital elements. In addition,
it makes possible to determine the orbital elements from the
precise speckle measurements only made with a large telescope
(4m).

 In our calculations of the orbital elements the
Kovalski-Olevi\'c method (Olevi\'c and Cvetkovi\'c 2004) is
applied. Due to the short arc covered by the speckle measurements
we obtain several good fits with periods of 62-75 years and our
final orbit (Table 1) is chosen only after fitting the radial
velocity measurements to our orbital elements (Fig. 2).

The orbital elements (equinox J2000) are listed in Table 1
together with the orbital elements following from the two earlier
solutions: first orbit (Gies et al. 1993) and second orbit (Gies
et al.). Our period and semimajor axis are 3.1 and 2.6 times
longer than the corresponding values of the preceding solution,
respectively. The inclination is twice the previous one, but in
the case of eccentricity and epoch of periastron similar values
are obtained.

Fig. 1 gives the apparent orbit where the solid curve refers to
our orbit and the dashed one to that published earlier. The solid
straight line is the nodal line. The arrow in the corner below indicates the sense
of the companion's revolution around the primary. From the figure
it is clearly seen that the arcs of both orbits almost coincide
between the moments of the first measurement and the one about the
periastron, to differ significantly afterwards.

\begin{figure*}
\centerline{\includegraphics[height=8.5cm]{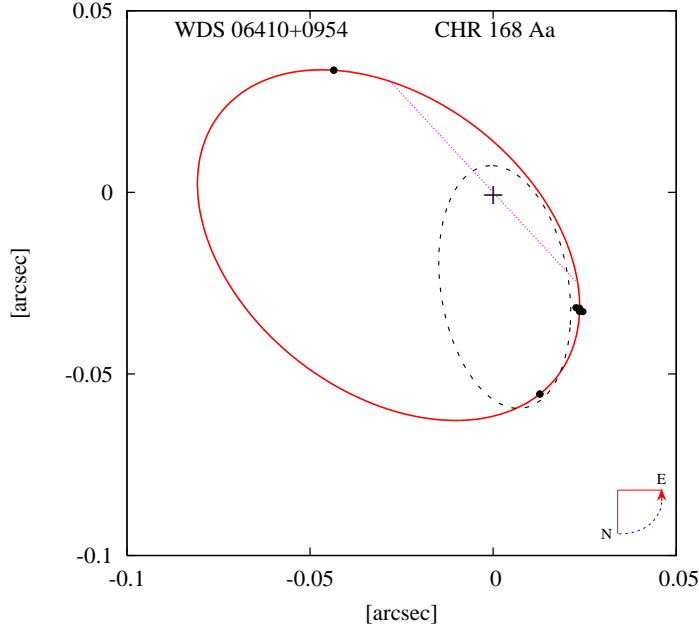}}
\caption{Apparent orbit for 15 Monocerotis: the solid curve refers
to our orbit and the dashed one to that published earlier (second
orbit).}
\end{figure*}

Table 2 contains the observational data used and their residuals
for all three calculated orbits.

Finally, Table 3 gives the ephemerides for the interval 2008-2012.

\begin{table*}
 \centering
  \caption{Orbital elements for 15 Monocerotis = CHR 168 Aa  }\label{T1}
  \begin{tabular}{ccccccccccc} \hline\hline
Orbit       &  $P$[yr]   & $T$      & $a[^{\prime\prime}]$ & $e$  & $i[^\circ]$ & $\Omega[^\circ]$ & $\omega[^\circ]$ \\
\hline
            &             &            &             &             &            &           &          \\
First orbit &     25.2    &    1922.75 &      0.0339 &      0.67   &    30.4    &     15.0  &    169.  \\
            &       -     &      -     &  $\pm$0.015 &       -     &  $\pm$10.0 &  $\pm$2.5 &     -    \\
            &             &            &             &             &            &           &          \\
Second orbit&     23.6    &    1926.0  &      0.0339 &      0.78   &    35.     &     16.8  &    168.  \\
            & $\pm$12.    &  $\pm$11.  &  $\pm$0.030 &  $\pm$0.3   &   $\pm$5.  &  $\pm$100.& $\pm$45. \\
            &             &            &             &             &            &           &          \\
Our orbit   &    74.00    &    1996.07 &      0.0885 &      0.76  &    62.4    &     42.6  &    82.6  \\
            & $\pm$0.30   &  $\pm$0.29 &  $\pm$0.0028&  $\pm$0.017 &   $\pm$0.4 &  $\pm$0.4 & $\pm$1.5 \\
\hline\hline
\end{tabular}
\end{table*}

\begin{table*}
 \centering
  \caption{Residuals  }\label{T1}
  \begin{tabular}{ccccccc} \hline\hline
    &  &   &  &  \ \ first orbit \ \ \ & \ \ second orbit \ \ \ & \ \ our orbit \ \ \ \\
Epoch  & $\theta$ & $\rho$  & Reference  & (O-C)$_{\theta}$ \ (O-C)$_{\rho}$  & (O-C)$_{\theta}$ \ (O-C)$_{\rho}$ & (O-C)$_{\theta}$ \ (O-C)$_{\rho}$   \\
\hline
   1988.1704 &   12.9  &  0.057   &  McA1993  &   $-$0.6  \ \  0.002 &     $-$2.1  \  0.000 &    0.0  \ \  0.000 \\
   1993.0925 &   35.4  &  0.039   &  McA1993  &     \ 0.3 \ $-$0.002 &        1.5  \ \ 0.001 & $-$0.3 \ $-$0.001 \\
   1993.1967 &   36.7  &  0.041   &  McA1993  &      1.0  \ \ \ 0.000 &        2.2 \ \  0.004 &    0.3 \ \   0.001 \\
   2001.0197 &  232.3  &  0.055   &  Msn2004  &  $-$74.5  \ \ 0.030 &   $-$108.9   \ 0.016 &    0.0   \ \ 0.000 \\
\hline\hline
\end{tabular}
\end{table*}

\begin{table*}
 \centering
  \caption{Ephemerides }\label{T3}
  \begin{tabular}{ccccc}
  \hline\hline
     2008 & 2009 & 2010 & 2011 & 2012 \\
%          &      &      &      &      \\
 $\theta $ \ \quad $\rho $ &
 $\theta $ \ \quad $\rho $ \ \ &
 $\theta $ \ \quad $\rho $ &
 $\theta $ \ \quad $\rho $ &
 $\theta $ \ \quad $\rho $ \\
 $[^o] $ \ $[^{\prime\prime}]$ &
 $[^o] $ \ $[^{\prime\prime}]$ \ \ &
 $[^o] $ \ $[^{\prime\prime}]$ &
 $[^o] $ \ $[^{\prime\prime}]$ &
 $[^o] $ \ $[^{\prime\prime}]$ \\
\hline
     &      &      &      &      \\
 250.3 \ \ 0.076& 252.3 \ \ 0.077 & 254.2 \ \ 0.078 & 256.1 \ \ 0.079 & 257.9\ \ 0.079\\
\hline \hline
\end{tabular}
\end{table*}

\section{Spectroscopic Data}

Spectroscopic observations of 15 Mon are unequally spaced in time
with large gaps (the longest gap is 46 years). Because of this the
orbital elements of this binary determined from radial velocities
are rather uncertain. It has to be emphasized that some very good,
high resolution spectroscopic observations (S/N$\approx $400),
indicate the presence of some spectral features in the wings of
rather sharp and deep primary line profiles that could be formed
by spectral lines which belong to the secondary component (Gies et
al 1993). These very shallow and broad features, blended by strong
spectral lines of the primary star, are not suitable for precise
radial velocity measurements of the secondary component. Moreover,
possible weak emission components of these lines may cause a
slight distortion of line profiles inducing more problems in the
line position measurements. Therefore, practically 15 Mon is a
single-lined binary system. Consequently, from radial velocity
measurements, in addition to the eccentricity and the periastron
longitude, the mass function and the projected semimajor axis of
the primary star can be derived only.

For orbital elements determination the radial velocity data have
to be placed into the phase order for which the orbital period is
required. Since 15 Mon is not an eclipsing binary, the orbital
period can not be derived from light-curve observations, but only
from the radial velocity data or/and from speckle determinations
of orbital elements. This method of orbital period determination
of 15 Mon was employed by Gies et al. (1993, 1997). There is more
or less difference between old (Gies et al. 1993) and new (Gies et
al. 1997) orbital elements. For instance, the new orbital period
(23.6 years) is slightly shorter than the old one, but the mass
function and the projected semimajor axis are about twice larger
for the new orbit than for the old one. These discrepancies
introduce difficulties in the interpretation of the physical
parameters of the binary components.

Combining our orbital period, eccentricity and inclination of the
orbit, and the new value of the semiamplitude in the velocity
curve given by Gies et al. (1997) we obtain 3.812 for the mass
function. We estimate the mass of the primary to be 29.1${\cal
M}_{\odot}$, based on its position in the HR diagram according to
spectral type O7V (Lang 1992) and derive the mass of the secondary
 21.3${\cal M}_{\odot}$. This leads to a mass ratio of 0.73. The
obtained mass ratio is consistent with the expected mass ratio
derived from the spectral type of the components. Thus the total
mass is 50.4${\cal M}_{\odot}$. Combined with our orbital elements
by means of Kepler's third law it yields a dynamical distance of
736 pc which is within the distance interval mentioned in
Introduction. Therefore, there is no need to assume the
unrealistic distance of 950 pc (Gies et al. 1993).

With our orbital elements we calculate the radial velocity in
order to see its agreement with the measured values. This is
presented in Fig. 2. The best agreement is found when values of
about 28-30 km s$^{-1}$ are assumed for the motion of the mass
center along the line of sight. Independently we find 28.4 km/s in
two catalogs (Barbier-Brossat \& Figon 2000, Kharchenko et al.
2007). Therefore, this value is assumed as the radial velocity of
the mass center in the fitting procedure. The fit in the radial
velocity helped assume the values given for our orbital elements
in Table 1 as final ones since we had several good solutions as
said above. In this case the residuals between the values of
radial velocity following from our assumed orbit and the measured
values are the smallest.

\begin{figure*}
\centerline{\includegraphics[height=8.5cm]{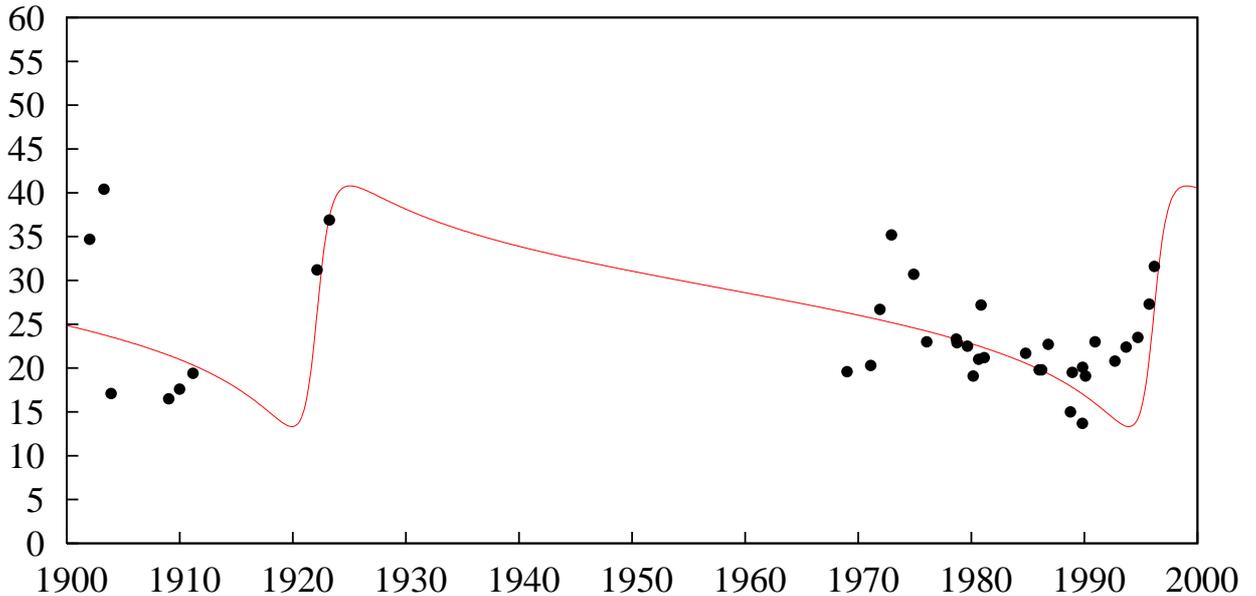}}
\caption{Radial velocity fit: the curve corresponds to our orbital
elements and the points are the measured values.}
\end{figure*}

\section{Photometric Distance}

For 15 Mon the distance can be determined also photometrically.
The data about the spectral type are available, especially for the
primary (O7V) where they are very reliable, whereas for the other
component we find O9.5V. Then the absolute magnitude $M_v $
expected for such spectral type in the case of component A is
equal to $-$5.2 (e.g. Lang 1992 - p.137, Binney \& Merrifield 1998
p.110; both based on the same data source from 1982). Combined
with the value for the corresponding apparent magnitude (4.66)
this yields a distance modulus of 9.86. This value means that the
binary would be at a distance of about 950 pc, but provided that
the interstellar extinction is zero which, clearly, is not
realistic.

However, for O stars in general we find physical parameters in
(Howarth \& Prinja 1989) according to which an O7V star should
have $M_v = -4.9$. This yields a distance modulus of 9.56, i.e.
without extinction the distance would be about 815 pc.

For the interstellar extinction we find various values: from about
0.2 (Lang 1992, Dahm \& Simon 2005) via 0.25 (Dahm et al. 2007)
towards 0.41 (Ram\'irez et al. 2004). With regard to these values
most likely the distance of the binary lies between 700 pc and 800
pc. This result agrees very well with the values cited above for
cluster NGC 2264 as a whole.

\section{Discussion and Conclusions}

If accepted that the limits for the 15 Mon distance are 700 pc and
800 pc (more precisely the middle of 750 pc is assumed) and the
orbital elements proposed by Gies et al. (1997) are used, then by
applying Kepler's third law one obtains the corresponding total
mass to be equal to 29.5${\cal M}_{\odot}$. If our orbital
elements are used, by applying Kepler's third law the total mass
of the system is found to be 53.4${\cal M}_{\odot}$ for the same
distance.

Among the physical parameters of O stars determined by Howarth \&
Prinja (1989) is also the mass. According to them an O7V star
should have a mass of 36${\cal M}_{\odot}$ and an O9.5V one
21${\cal M}_{\odot}$. In the particular case the primary of 15 Mon
was found by these two authors to have a mass of 39${\cal
M}_{\odot}$. Then the total mass of this binary is 60${\cal
M}_{\odot}$. Our total mass of 53.4${\cal M}_{\odot}$ is very
close to this value and the one following from the spectroscopy
(50.4${\cal M}_{\odot}$) is not too different. However, the value
of 29.5${\cal M}_{\odot}$ following from the orbital elements by
Gies et al. (1997) appears to be too low. In addition, the
mass-luminosity relation for the main sequence stars (e.g. Lang
1992) yields for the total mass of such a pair about 50${\cal
M}_{\odot}$.

From what has been said above we conclude that:

i) the distance of 950 pc for 15 Mon (open cluster NGC 2264) is too large, most
likely the true one lies between 700 pc and 800 pc;

ii) our orbital elements obtained from speckle measurements only are in
favor of a longer period and larger semimajor axis than found
earlier.

New measurements both speckle and radial velocity ones are very
desirable in order to throw more light on the case of binary 15
Mon.

 \section*{Acknowledgments}

 This research has been supported by the Ministry of Science and
Environmental Protection of the Republic of Serbia (Project No
146004 "Dynamics of Celestial Bodies, Systems and Populations").

\section*{References}

{Barbier-Brossat M. \& Figon P.} {2000} {Astron. Astrophys. Suppl.
Ser.} {142} {217}

{Binney, J. and Merrifield, M.} {1998} {Galactic Astronomy}
{Princeton University Press, Princeton, New Jersey, p.110}

{Dahm, S.E. \& Simon, T.} {2005} {AJ} {129} {829}

{Dahm, S.E., Simon, T., Proszkow, E.M. \& Patten, B.M.} {2007} {AJ}
{134} {999}

{ESA} {1997} {The Hipparcos and Tycho Catalogues} {ESA SP-1200}

{Gies, D.R., Mason, B.D., Hartkopf, W.I., McAlister, H.A., Frazin,
R.A., Hahula, M.E., Penny, L.R. \& Thaller, M.L.} {1993} {AJ} {106}
{2072}

{Gies, D.R., Mason, B.D., Bagnuolo, Jr.W.G., Hahula, M.E., Hartkopf,
W.I., McAlister, H.A. \& Thaller, M.L.} {1997} {ApJ} {475} {L49}

{Hartkopf, W.I. and Mason, B.D.} {2006} {Sixth Catalog of Orbits of
Visual Binary Stars} {US Naval Observatory, Washington} {Electronic
version http://ad.usno.navy.mil/wds/orb6.html}

{Hartkopf, W.I., Mason, B.D., Wycoff, G.I. \& McAlister, H.A.}
{2006} {Fourth Catalog of Interferometric Measurements of Binary
Stars}, {US Naval Observatory, Washington} {Electronic version
http://ad.usno.navy.mil/wds/int4.html}

{Howarth, I.D. \& Prinja, R.K.} {1989} {ApJS} {69} {527}

 {Kharchenko N.V., Scholz R.-D., Piskunov A.E., Roeser S.
 \& Schilbach E.} (2007) {Astron. Nachr.} {328} {889}

{Lang, K.R.} {1992} {Astrophysical Data: Planets and Stars,
Springer-Verlag}

{Mason, B.D., Wycoff, G.L. and Hartkopf, W.I.} {2006} {The
Washington Visual Double Star Catalogue} {US Naval Observatory,
Washington} {Electronic version
http://ad.usno.navy.mil/wds/wds.html}

{Olevi\'c, D. \& Cvetkovi\'c, Z.} {2004} {A\&A} {425} {25..}

{P\'erez, M.R., Th\'e P.S. \& Westerlund, B.E.} {1987} {PASP} {99}
{1050}

{Pourbaix, D. et al.} {2004} {The ninth catalogue of spectroscopic
binary orbits} {Electronic version http://sb9.astro.ulb.ac.be}

{Ram\'irez, S.V., Rebull, L., Stauffer, J., Hearty, T., Hillenbrand,
L., Jones, B., Makidon, R., Pravdo, S., Strom, S. \& Werner, M.}
{2004} {AJ} {127} {2659}

\end{document}